\documentclass[aps,prb,twocolumn,amssymb,amsmath,superscriptaddress]{revtex4-1}

\usepackage{bm}
\usepackage{color}
\usepackage{graphicx}
\usepackage{dcolumn}
\usepackage{graphicx}
\usepackage[colorlinks=true, letterpaper=true, pdfstartview=FitV, linkcolor=blue, citecolor=blue, urlcolor=blue]{hyperref}
\usepackage[normalem]{ulem}
\usepackage{amsmath}
\usepackage{epstopdf}
\usepackage{multirow}
\usepackage{slashed}
\usepackage{float}
\usepackage{pifont}

\setcitestyle{square}

\makeatletter
\renewcommand\@biblabel[1]{#1.}
\makeatother

\begin{document}
	
	\title{Intrinsic and extrinsic anomalous transport properties in noncollinear antiferromagnetic Mn$_3$Sn from first-principle calculations}
	
	\author{Xiuxian Yang}
	\affiliation{Key Lab of advanced optoelectronic quantum architecture and measurement (MOE), Beijing Key Lab of Nanophotonics $\&$ Ultrafine Optoelectronic Systems, and School of Physics, Beijing Institute of Technology, Beijing 100081, China}
	\affiliation{School of Physics, Jiangsu Normal University, Xuzhou 221116, China}
	
	\author{Wanxiang Feng}
	\email{wxfeng@bit.edu.cn}
	\affiliation{Key Lab of advanced optoelectronic quantum architecture and measurement (MOE), Beijing Key Lab of Nanophotonics $\&$ Ultrafine Optoelectronic Systems, and School of Physics, Beijing Institute of Technology, Beijing 100081, China}
	
	\author{Xiaodong Zhou}
	\affiliation{Laboratory of Quantum Functional Materials Design and Application, School of Physics and Electronic Engineering, Jiangsu Normal University, Xuzhou 221116, China}
	
	\author{Yuriy Mokrousov}
	\affiliation{Institute of Physics, Johannes Gutenberg University Mainz, 55099 Mainz, Germany}
	\affiliation{Peter Gr\"unberg Institut and Institute for Advanced Simulation, Forschungszentrum J\"ulich and JARA, 52425 J\"ulich, Germany}
	
	\author{Yugui Yao}
	\email{ygyao@bit.edu.cn}
	\affiliation{Key Lab of advanced optoelectronic quantum architecture and measurement (MOE), Beijing Key Lab of Nanophotonics $\&$ Ultrafine Optoelectronic Systems, and School of Physics, Beijing Institute of Technology, Beijing 100081, China}
	
	\date{\today}
	
	\begin{abstract}
		Mn$_3$Sn has garnered significant attention due to its kagome lattice, 120$^\circ$ noncollinear antiferromagnetic order, and substantial anomalous Hall effect. In this study, we comprehensively explore intrinsic and extrinsic contributions to anomalous Hall, anomalous Nernst, and anomalous thermal Hall effects, employing first-principle calculations and group theory analysis.  Comparative analysis between our theoretical results and available experimental data underscores the predominance of intrinsic mechanism in shaping anomalous transport properties at low temperatures.  Specifically, Weyl fermions are identified as the primary contributors to intrinsic anomalous Hall conductivity. The significance of extrinsic mechanisms becomes evident at high temperatures, especially when the longitudinal charge conductivity falls into the dirty regime, where the side jump mechanism plays a vital role.  Extrinsic contributions to anomalous transport properties are primarily influenced by the electronic states residing at the Fermi surfaces.  Furthermore, anomalous transport properties exhibit periodic variations when subjected to spin rotations within the kagome plane, achievable by applying an external magnetic field.  Our findings advance the understanding of anomalous transport phenomena in Mn$_3$Sn and offer insights into potential applications of noncollinear antiferromagnetic materials in spintronics and spin caloritronics.
	\end{abstract}
	
	\maketitle
	
	\section{Introduction}\label{intro}
	
	The anomalous Hall effect (AHE), discovered by Hall in 1881, refers to the emergence of a transverse charge current in response to a longitudinal electric field without the presence of an external magnetic field~\cite{Hall1881}. It remains a fundamental aspect of condensed matter physics, shedding light on the intricate nature of magnetism~\cite{Nagaosa2010}. Over time, the understanding of the physical mechanisms underlying AHE has evolved, dividing the effect into intrinsic and extrinsic components. The intrinsic mechanism, which is not influenced by electron scattering, was initially proposed by Karplus and Luttinger~\cite{Karplus1954}, and is now well explained by Berry phase theory, relying solely on the electronic band structure of pristine crystals~\cite{Sundaram1999,Yao2004}. In contrast, the extrinsic mechanisms, such as skew scattering~\cite{Smit1955,Smit1958} and side jump~\cite{Berger1970}, hinge on electron scattering caused by impurities or disorder.  Moreover, there are two other remarkable anomalous transport phenomena: the anomalous Nernst effect (ANE)\cite{XiaoD2006} and the anomalous thermal Hall effect (ATHE)\cite{Qin2011}, which involve the emergence of transverse charge and heat currents driven by longitudinal temperature gradients, respectively.  Analogous to the anomalous Hall conductivity (AHC) $\sigma_{ij}$, the anomalous Nernst and anomalous thermal Hall conductivities (ANC and ATHC), $\alpha_{ij}$ and $\kappa_{ij}$, can be decomposed into three distinct parts:
	\begin{eqnarray}
		\sigma_{ij} & = &\sigma_{ij}^{\textnormal{int}}+\sigma_{ij}^{\textnormal{sj}}+\sigma_{ij}^{\textnormal{isk}}, \label{eq:sigma-Decomposed} \\
		\alpha_{ij} & = &\alpha_{ij}^{\textnormal{int}}+\alpha_{ij}^{\textnormal{sj}}+\alpha_{ij}^{\textnormal{isk}},\label{eq:alpha-Decomposed}\\
		\kappa_{ij} & = &\kappa_{ij}^{\textnormal{int}}+\kappa_{ij}^{\textnormal{sj}}+\kappa_{ij}^{\textnormal{isk}}, \label{eq:kappa-Decomposed}
	\end{eqnarray}
	Here, the subscripts ${i,j}\in{x,y,z}$ represent Cartesian coordinates, and the superscripts ${\textnormal{int}}$, ${\textnormal{sj}}$, and ${\textnormal{isk}}$ denote the intrinsic, side jump, and skew scattering contributions, respectively.
	
	The AHE is commonly observed in ferromagnetic conductors and it is assumed to be proportional to the magnetization. In contrast, antiferromagnets (AFMs) have long been considered to lack the AHE due to their zero net magnetization~\cite{Jungwirth2002,Fang2003,Onoda2006,Onoda2008,Onose_Lor2008,Shiomi_Lor2009,Shiomi_Lor2010,Zhou2019}. However, recent advancements have challenged this notion. For example, a significant AHC was predicted in the noncollinear antiferromagnetic Mn$_3$Ir through a combination of symmetry analysis and first-principles calculations~\cite{Chen2014}. Subsequent experiments have confirmed substantial AHC in noncollinear AFMs Mn$_3$\emph{X} (\emph{X} = Sn, Ge) even in the absence of an external magnetic field~\cite{Nakatsuji2015,Naoki2016,Nayak2016}.  Compared to ferromagnets, AFMs exhibit an array of exotic properties, including insensitivity to magnetic-field perturbations~\cite{Marti2014,Wadley2016}, ultrafast spin dynamics~\cite{Andrei2010}, and high-frequency uniform spin precession~\cite{Tzschaschel2017,Kampfrath2011,Pashkin_2013}. These attributes position AFMs as an excellent platform for antiferromagnetic spintronics~\cite{Jungwirth2016,Baltz2018}.
	
	Mn$_3$\emph{X}, as a representative family of noncollinear AFMs, has garnered significant attention due to its intriguing features, including substantial AHE~\cite{Nakatsuji2015,Naoki2016,Nayak2016,Matsuda2020,LC-Xu2020}, ANE~\cite{Muhammad2017,LiXK2017}, magneto-optical effects~\cite{Tomoya2018,HC-Zhao2021}, magnetic Weyl fermions~\cite{Kuroda2017,Chen2021}, magnetic spin Hall effect~\cite{Kimata2019}, and spin–orbit torque~\cite{Takeuchi2021,Higo2022}.  Moreover, Mn$_3$\emph{X} possesses a unique breathing-type kagome lattice structure formed by Mn atoms, as shown in Fig.~\ref{fig:structure}. This lattice hosts intriguing topological electronic bands, superconducting phases, and strong electromagnetic and transport responses\cite{Linda2018,Liu_2019,YinJX2020}, making it an ideal platform for exploring novel states of quantum matter.  However, previous theoretical investigations on Mn$_3$\emph{X}~\cite{Chen2014,Kubler_2014,Suzuki_2017,GuoGY2017,ZhangYang2017} have primarily focused on the anomalous transport properties induced by intrinsic Berry curvature mechanism, with limited attention paid to the extrinsic mechanisms related to the scattering of electrons off impurities or disorder.  In reality, understanding the contribution of extrinsic mechanisms to the AHE in kagome materials is crucial.  For instance, remarkable AHC values ranging from $10^4$ to $10^5$ S/cm, driven by extrinsic mechanisms, have been discovered in other kagome materials such as KV$_3$Sb$_5$~\cite{YangSY2020}, CsV$_3$Sb$_5$~\cite{YuFH2021}, and Nd$_3$Al~\cite{Singh2021}. These observations underscore the predominant role played by extrinsic mechanisms in governing the AHE, ANE, and ATHE in the kagome antiferromagnetic materials.
	
	In this work, we conduct a comprehensive investigation on the intrinsic and extrinsic mechanisms of anomalous transport properties, including the AHE, ANE, and ATHE, in noncollinear antiferromagnetic Mn$_3$Sn, using the state-of-the-art first-principles calculations.  By collectively rotating all spins within the kagome plane, we discern the tensor shapes of AHC, ANC, and ATHC through magnetic group theory. For nonzero tensor elements, we compute the intrinsic, side jump, and skew scattering contributions individually.  A profound anisotropy in the AHE, intricately connected to the evolving coplanar noncollinear spin configurations, is unveiled.  Through careful comparisons with available experimental data, we establish the consistent prevalence of the intrinsic mechanism in driving the AHE at low temperatures, notably when the longitudinal conductivity exceeds $10^4$ S/cm.  Our study highlights the influential role of Weyl fermions near the Fermi energy in shaping the intrinsic AHC in Mn$_3$Sn.  Nevertheless, we also observe a significant increase in the impact of extrinsic mechanisms, especially the side jump component, as the longitudinal conductivity falls below $10^4$ S/cm.  The extrinsic AHC predominantly emanates from electronic states positioned precisely at the Fermi surface sheets.  Furthermore, our calculations of ANC and ATHC, as well as the anomalous Lorentz ratio, consistently align with experimental observations at low temperatures. Through these findings, we advance the understanding of the intricate competition between intrinsic and extrinsic mechanisms that govern anomalous transport phenomena in the realm of noncollinear antiferromagnetic Mn$_3$Sn.

	\section{Theory and computational details}\label{method}
	
	The AHE, ANE, and ATHE are interconnected through the generalized Landauer-B\"uttiker formalism~\cite{ashcroft1976solid,Houten1992,behnia2015fundamentals} as expressed by the anomalous transport coefficients :
	\begin{equation}\label{eq:LB}
		R^{(n)}_{ij}=\int^\infty_{-\infty}(E-\mu)^n\left(-\frac{\partial f}{\partial E}\right)\sigma_{ij}(E)dE,
	\end{equation}
	where $\mu$ is the chemical potential, $f = 1/[\textnormal{exp}((E-\mu)/k_{B}T) + 1]$ represents the Fermi-Dirac distribution function, and $\sigma_{ij}$ is the AHC at zero temperature. The temperature-dependent ANC and ATHC can be expressed as follows:
	\begin{eqnarray}
		\alpha_{ij} & = &-R^{(1)}_{ij}/eT, \label{eq:ANC} \\
		\kappa_{ij} & = & R^{(2)}_{ij}/e^2T, \label{eq:ATHC}
	\end{eqnarray}
	From the equations \eqref{eq:LB} to \eqref{eq:ATHC}, it is evident that the AHC $\sigma_{ij}$ plays a crucial role in determining the other anomalous transport properties.
	
	Following the Kubo formalism within the linear-response theory~\cite{Kubo1957}, the AHC can be partitioned into Fermi surface ($\sigma_{ij}^\textnormal{I}$) and Fermi sea ($\sigma_{ij}^\textnormal{II}$) components~\cite{Czaja2014}:
	\begin{eqnarray}\label{eq:int_1}
		\sigma_{ij}^{\textnormal{I}}&=&-\frac{e^2\hbar}{2\pi}\int\frac{\textnormal{d}^3k}{2\pi^3}\sum_{m\neq n}{\rm Im}[v_{mn}^i(\textbf{k})v_{nm}^j(\textbf{k})] \nonumber\\
		&=&\frac{(E_{m\textbf{k}}-E_{n\textbf{k}})\Gamma}{[(E_f-E_{m\textbf{k}})^2+\Gamma^2][(E_f-E_{n\textbf{k}})^2+\Gamma^2]},
	\end{eqnarray}
	and
	\begin{eqnarray}\label{eq:int_2}
		\sigma_{ij}^{\textnormal{II}}&=&\frac{e^2\hbar}{\pi}\int\frac{\textnormal{d}^3k}{(2\pi)^3}\sum_{m\neq n}{\rm Im}[v_{mn}^i(\textbf{k})v_{nm}^j(\textbf{k})] \nonumber\\
		&=&\left\{\frac{\Gamma}{(E_{m\textbf{k}}-E_{n\textbf{k}})[(E_f-E_{m\textbf{k}})^2+\Gamma^2]}\right.  \nonumber\\
		&&\left. -\frac{1}{(E_{m\textbf{k}}-E_{n\textbf{k}})^2}{\rm Im}\left[{\rm In}\frac{E_f-E_m\textbf{k}+\textbf{i}\Gamma}{E_f-E_n\textbf{k}+\textbf{i}\Gamma}\right]\right\},
	\end{eqnarray}
	where $i,j\in x,y,z$ represent Cartesian coordinates, $v$ is the velocity operator, $E_f$ is the Fermi energy, $E_{n\textbf{k}}$ is the energy eigenvalue with band index $n$ at momentum $\textbf{k}$, and $\Gamma$ is an adjustable smearing parameter (0 $\sim$ 0.09 eV), respectively.  This constitutes the constant smearing (CS) model, which describes the intrinsic AHE. In this model, a constant $\Gamma$ parameter is assigned, providing all electronic states with the same finite lifetime. In the clean limit (i.e., $\Gamma\rightarrow 0$), the summation of Eqs.~\eqref{eq:int_1} and~\eqref{eq:int_2} converges to the well-established Berry curvature expression~\cite{Yao2004}:
	\begin{equation}\label{eq:IAHC} 
		\sigma^\textnormal{int}_{ij} = e^{2}\hbar \int\frac{\textnormal{d}^{3}k}{(2\pi)^{3}}\sum^\textnormal{occ}_{n,m \neq n}\frac{2\mathrm{Im}[v^i_{mn}(\textbf{k})v^j_{nm}(\textbf{k})]}{\left(E_{m\textbf{k}}-E_{n\textbf{k}}\right)^{2}}.
	\end{equation}
	It should be noted that the complex scattering mechanisms are not explicitly considered within the CS model.
	
	Alternatively, the inclusion of a short-range Gaussian disorder potential allows for the consideration of scattering-dependent AHC, encompassing the side jump and skew scattering mechanisms. Within the Gaussian disorder (GD) model, the impurity potential is described as
	\begin{equation}\label{eq:potential}
		V=U\sum_{i}^{N}\delta(\textbf{r}-\textbf{R}_{i}),
	\end{equation}
	where $U$ signifies the scattering strength, $\delta$ is the delta function, and $\textbf{R}_{i}$ corresponds to the $i$-th random atomic position among a total of $N$ impurities. Consequently, the impurity concentration is denoted as $n_\textnormal{i} = N/V$, with $V$ being the volume of the cell. For convenience, the disorder parameter is expressed as $\mathcal{V}=U^2n_\textnormal{i}$ ($0\sim80\ \textnormal{eV}^2a_0^3$).  It is crucial to emphasize that this impurity potential is spin-independent as it does not encompass spin degrees of freedom. With the incorporation of spin-orbit coupling, the electron's spin becomes intricately reliant on the modification of its orbital angular momentum during scattering. Although the impurity potential utilized in this context can only be interpreted as nonmagnetic impurities in magnetic materials, the possibility of a transverse flow of spin-polarized electrons induced by scattering (i.e., extrinsic anomalous Hall conductivity) is feasible, as demonstrated in previous works~\cite{Sinitsyn2007,Sinitsyn2008}. 
	
	The self-energy $\Sigma(E,\textbf{k})$, which accounts for the impact of electron scattering off impurities, can be expressed as follows, truncated to the lowest order~\cite{Czaja2014}:
	\begin{equation}\label{eq:Sigma}
		\Sigma(E,\textbf{k})=\mathcal{V}\int\frac{\textnormal{d}^3k'}{(2\pi)^3}O_{\textbf{kk}'}G_0(E,\textbf{k}')O_{\textbf{k}'\textbf{k}}.
	\end{equation}
	Here, $O_{\textbf{kk}'}$ represents the overlap matrix for the eigenstates at different momenta, and $G_0(E,\textbf{k}')=[E-H(\textbf{k}')]^{-1}$ stands for the bare Green's function with the unperturbed Hamiltonian $H(\textbf{k}')$.
	
	After accounting for the scattering effects, the AHC can be formulated using the full Green's functions $G^{R/A}$ ($R$: retarded and $A$: advanced)~\cite{Czaja2014} as follows:
	\begin{eqnarray}\label{eq:ext_1}
		\sigma_{ij}^\textnormal{I}&=&\frac{e^2\hbar}{4\pi}\int\frac{\textnormal{d}^3k}{(2\pi)^3}{\rm Tr}[\Gamma^i (E_f, \textbf{k})G^R(E_f, \textbf{k})v^jG^A(E_f, \textbf{k}) \nonumber\\
		&&-(i\leftrightarrow j)],
	\end{eqnarray}
	and
	\begin{eqnarray}\label{eq:ext_2}
		\sigma_{ij}^\textnormal{II}&=&\frac{e^2\hbar}{2\pi}\int\frac{\textnormal{d}^3k}{(2\pi)^3}\int^{E_f}_{-\infty}{\rm Re}\lbrace{\rm Tr}[\Gamma^i(E, \textbf{k})G^R(E, \textbf{k})  \nonumber\\
		& &\times \gamma(E, \textbf{k}) G^R(E, \textbf{k})\Gamma^j(E, \textbf{k})G^R(E, \textbf{k})  \nonumber\\
		& &-(i\leftrightarrow j)]\rbrace \textnormal{d}E.
	\end{eqnarray}
	Here, $\gamma(E, \textbf{k})$ and $\boldsymbol{\Gamma}(E, \textbf{k})$ are scalar and vector vertex functions, respectively, defined as
	\begin{eqnarray}\label{eq:gamma}
		\gamma(E, \textbf{k})&=&I+\mathcal{V}\int\frac{\textnormal{d}^3k'}{(2\pi)^3}O_{\textbf{kk}'}G^R(E, \textbf{k}')\gamma(E, \textbf{k}')\nonumber\\
		&&\times G^R(E,\textbf{k}')O_{\textbf{k}'\textbf{k}},
	\end{eqnarray}
	and
	\begin{eqnarray}\label{eq:Gamma}
		\boldsymbol{\Gamma}(E, \textbf{k})&=&\boldsymbol{v}(\textbf{k})+\mathcal{V}\int\frac{\textnormal{d}^3k'}{(2\pi)^3}O_{\textbf{kk}'}G^A(E, \textbf{k}')\boldsymbol{\Gamma}(E, \textbf{k}') \nonumber\\
		&&\times G^R(E, \textbf{k}')O_{\textbf{k}'\textbf{k}},
	\end{eqnarray}
	where $I$ and $\boldsymbol{v}$ are identity and velocity vector operators, respectively.  The Fermi sea term, Eq.~\eqref{eq:ext_2}, is conventionally regarded as intrinsic, devoid of any scattering-driven behavior.  In contrast, the Fermi surface term, Eq.~\eqref{eq:ext_1}, encompasses intrinsic, side jump, and skew scattering contributions.  Examining Eq.~\eqref{eq:ext_1}, if the bare Green function $G_0$ replaces the full Green function $G$ and the vertex correction is not considered (i.e., $\Gamma^i\rightarrow v^i$), it reflects an intrinsic contribution and yields intrinsic AHC ($\sigma_{ij}^{\textnormal{int}}$) when combined with the Fermi sea term.  When the full Green function $G$ is used and the vertex correction is not considered (i.e., $\Gamma^i\rightarrow v^i$), the side jump contribution to the AHC ($\sigma_{ij}^{\textnormal{sj}}$) emerges.  Finally, if the full Green function $G$ is used and the vertex correction is included (i.e., using $\Gamma^i$), the skew scattering contribution to the AHC ($\sigma_{ij}^{\textnormal{isk}}$) is introduced. The decomposition of AHC can be elucidated through Feynman diagrams, referring to Czaja et al.'s work~\cite{Czaja2014}. By plugging the decomposed AHC into Eqs.~\eqref{eq:LB}-\eqref{eq:ATHC}, the corresponding components of ANC and ATHC can be obtained accordingly.
	
	In the GD model, the skew scattering term is also known as ``intrinsic" skew scattering ($\sigma_{ij}^{\textnormal{isk}}$), originally proposed by Sinitsyn and co-workers.~\cite{Sinitsyn2007,Sinitsyn2008}.  Similar to conventional skew scattering, ``intrinsic" skew scattering also arises from the asymmetric part of the collision kernel. However, it converges to a finite value in the clean limit ($\mathcal{V}\rightarrow0$). In contrast, conventional skew scattering is inversely proportional to impurity concentrations and becomes divergent in the clean limit. Diagrammatically speaking, ``intrinsic" skew scattering solely results from Gaussian disorder correlations, while conventional skew scattering involves vertex corrections that include correlators of three or more disorder vertices~\cite{Czaja2014}.
	
	The Gaussian disorder model utilized in this study does not explicitly define the types (such as crystal defect or phonon) and spin structures of impurities. In adopting a ``mean-field" approach, the Gaussian disorder model accommodates various scattering channels without delving into the detailed characteristics of the internal nature of scattering sources. Taking into account temperature effects, the microscopic motions within the crystal become more intricate, potentially introducing variations between theoretical calculations and experimental measurements. A comprehensive disorder potential that encompasses all these details could offer a more accurate representation of electronic conductivity and its individual decomposed components. However, the computational treatment of these scattering processes at a detailed microscopic level remains a challenging task for first-principles methods.  Thus, the Gaussian disorder model proves suitable for Mn$_3$Sn, identified as a moderately disordered metal, given that its longitudinal conductivity falls within the dirty and intrinsic regimes ($\sigma_{ii}<10^6$ S/cm), but not the clean regime ($\sigma_{ii}>10^6$ S/cm), as illustrated in Fig.~\ref{fig:AHC-disorder}.
	
	The first-principle calculations are carried out using the full-potential linearized augmented plane-wave (FP-LAPW) method implemented in the \textsc{fleur} code~\cite{fleur}. The exchange-correlation functional is treated within the generalized gradient approximation using the Perdew-Burke-Ernzerhof parameterization~\cite{Perdew1996}.  The spin-orbit coupling is included in all calculations. For Mn$_3$Sn, a plane-wave cutoff energy of 3.80 a$_0^{-1}$ is selected, and the experimental lattice constants ($a = b = 5.66$ {\AA} and $c = 4.53$ {\AA}) are adopted. The self-consistent calculations and magnetic anisotropy energy calculations are conducted with a 16$\times$16$\times$18 mesh of $k$-points.  To construct maximally localized Wannier functions, $s$, $p$, and $d$ orbitals of Mn atoms, as well as $s$ and $p$ orbitals of Sn atoms, are projected onto a uniform $k$-mesh of 8$\times$8$\times$8 using the \textsc{wannier90} package~\cite{Pizzi_2020}. For calculating the AHC, an ultra-dense $k$-mesh of 300$\times$300$\times$300 is employed. For the calculations of the ANC and ATHC using Eq.~\eqref{eq:LB}, the AHC is computed with an energy interval of 0.1 meV.

	\section{Results and discussion}\label{results}
	
\begin{figure}[t]
		\includegraphics[width=1\columnwidth]{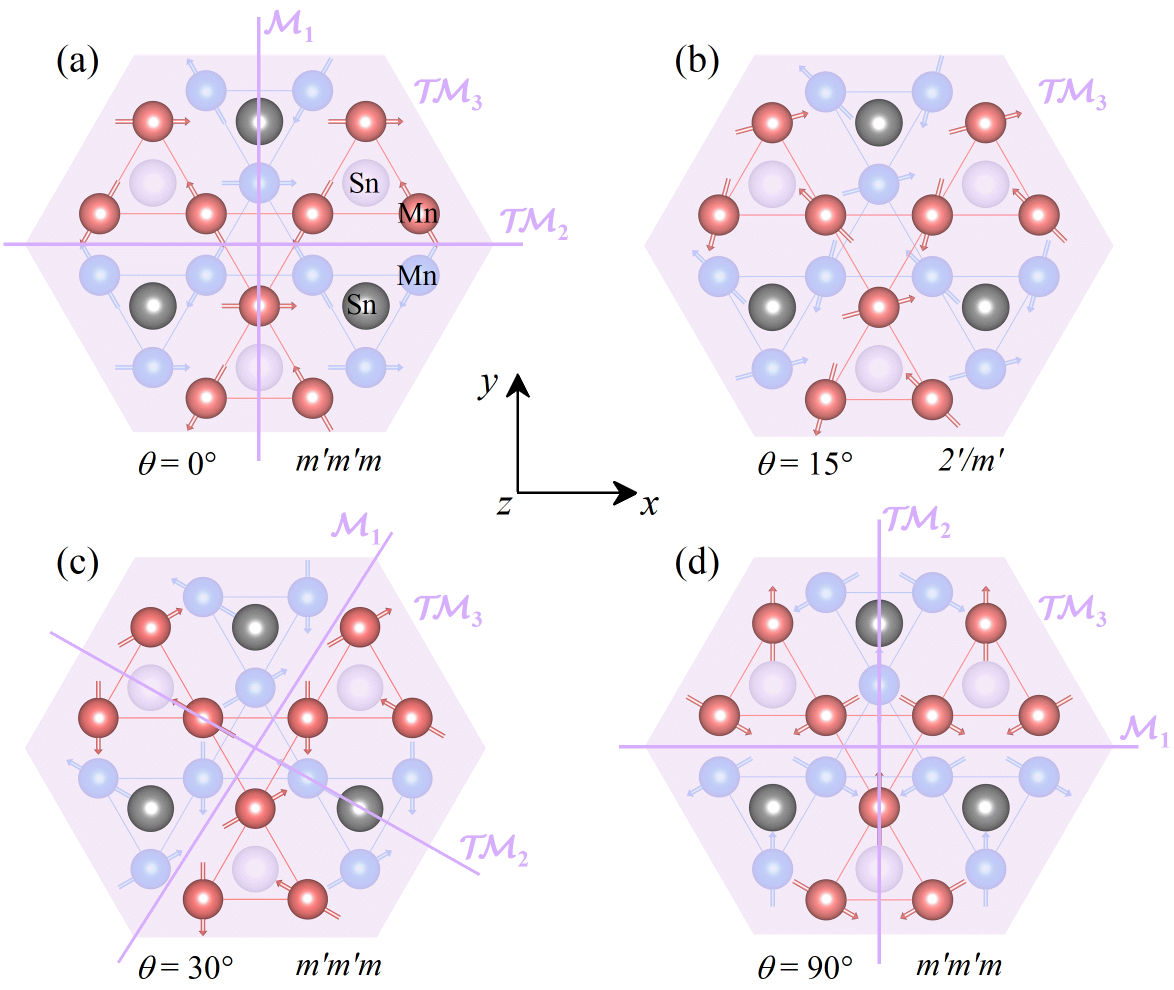}
		\caption{(Color online)  (a-d) The crystal and magnetic structure of noncollinear antiferromagnetic Mn$_3$Sn. The Mn atoms on the upper and lower kagome planes are marked in red and blue, respectively. Within the kagome plane, all spins can be uniformly rotated by an angle $\theta$.  The magnetic point groups at different $\theta$ are: (a) $m'm'm$ at $\theta = 0^\circ$, (b) $2'/m'$ at $\theta = 15^\circ$, (c) $m'm'm$ at $\theta = 30^\circ$, and (d) $m'm'm$ at $\theta = 90^\circ$. These spin configurations are characterized by specific symmetry elements: $\mathcal{M}_{1}$ denoting a mirror plane perpendicular to the kagome plane, $\mathcal{TM}_2$ representing a combined symmetry with the time-reversal symmetry $\mathcal{T}$ and a mirror plane $\mathcal{M}_2$ perpendicular to the kagome plane,  $\mathcal{TM}_3$ representing a combined symmetry with $\mathcal{T}$ and a mirror plane $\mathcal{M}_3$ parallel to the kagome plane.}
		\label{fig:structure}
	\end{figure}

	Bulk Mn$_3$Sn alloy crystallizes in a layered hexagonal structure with the crystallographic space group of $P6_3/mmc$. The primitive unit cell consists of two atomic layers stacked along the $c$ axis. Within each layer, the arrangement of three Mn atoms forms a kagome lattice, while the Sn atom is positioned at the center of each hexagon, as depicted in Fig.~\ref{fig:structure}. The spin magnetic moments of the three Mn atoms on the same kagome plane adopt a 120$^\circ$ noncollinear antiferromagnetic order with a N\'eel temperature ($T_N$) of 430 K~\cite{Nakatsuji2015,Muhammad2017}. Our calculated spin magnetic moment for each Mn atom is 3.26 $\mu_B$, which closely matches the experimental value of $\sim$ 3.0 $\mu_B$~\cite{Nakatsuji2015}.  Despite being classified as a noncollinear AFM, Mn$_3$Sn exhibits a very small net magnetic moment ($\sim$ 0.002 $\mu_B$)~\cite{Muhammad2017}. This residual magnetic moment allows for the manipulation of the spin orientation within the kagome plane, for instance, through an external magnetic field. Such spin rotations alter the magnetic group and total energy of the system, and consequently impact the anomalous transport properties.
	
	In this context, examining the variations in the AHC tensor due to spin rotation is adequate, since the ANC and ATHC share the same symmetry requirements according to Eqs.~\eqref{eq:LB}-\eqref{eq:ATHC}. The off-diagonal elements of the AHC can be represented in a vector notation as $\bm{\sigma} = [\sigma^x, \sigma^y, \sigma^z] = [\sigma_{yz}, \sigma_{zx}, \sigma_{xy}]$. Notably, the anomalous Hall vector $\bm{\sigma}$ can be analogously regarded as a pseudovector, akin to spin. Given that the translational operation ($\tau$) does not alter the anomalous Hall vector~\cite{Suzuki_2017}, i.e.,  $\tau\bm{\sigma}= \bm{\sigma}$, our subsequent analysis is effectively limited to magnetic point groups. This magnetic symmetry analysis has been previously employed in the study of other two- and three-dimensional magnetic materials~\cite{Zhou2019,XD-Zhou2019b,XD-Zhou2020,Zhang_2021,XD-Zhou2023}.  Table~\ref{tab:group} underscores that the magnetic point group of Mn$_3$Sn demonstrates a periodicity of $30^\circ$: $m'm'm$ $\rightarrow$ $2'/m'$ $\rightarrow$ $m'm'm$, as the spin rotates within the kagome plane ($x$-$y$ plane).  Here, it is only worth discussing two non-repetitive groups, namely $m'm'm$ and $2'/m'$, in relation to four distinct spin configurations ($\theta= 0^\circ$, $15^\circ$, $30^\circ$, and $90^\circ$), illustrated in Fig.~\ref{fig:structure}.  Note that the specific mirror symmetries plotted in Fig.~\ref{fig:structure} only characterize the spin configurations but not preserve the crystal structure.  First, for the $m'm'm$ group ($\theta= 0^\circ$), it consists of a mirror symmetry $\mathcal{M}_1$ and two combined symmetries $\mathcal{TM}_2$ and $\mathcal{TM}_3$, as depicted in Fig.~\ref{fig:structure}(a). The mirror plane $\mathcal{M}_1$ is perpendicular to the $x$-axis and parallel to the $y$-axis, which leads to a sign change in $\sigma^y$ and $\sigma^z$ while leaving $\sigma^x$ unchanged. Similarly, the mirror plane $\mathcal{M}_2$ (parallel to $x$-axis, perpendicular to $y$-axis) changes the signs of $\sigma^x$ and $\sigma^z$ but preserves $\sigma^y$.  As for the time-reversal symmetry altering signs of all $\sigma^{x}$, $\sigma^{y}$, and $\sigma^{z}$, the combined $\mathcal{TM}_2$ symmetry changes the sign of $\sigma^y$ but preserves $\sigma^x$ and $\sigma^z$.  Since $\mathcal{M}_3$ (perpendicular to $z$-axis, parallel to $x$-$y$ plane, between two kagome planes) changes the signs of $\sigma^x$ and $\sigma^y$, the combined symmetry $\mathcal{TM}_3$  preserves $\sigma^x$ and $\sigma^y$.   Consequently, for the $m'm'm$ group ($\theta= 0^\circ$), we find $\bm{\sigma} = [\sigma^x, 0, 0] = [\sigma_{yz}, 0, 0]$.  Second, when the spin rotates to 30$^\circ$ within the same $m'm'm$ group, the positions of $\mathcal{M}_1$ and $\mathcal{M}_2$ mirror planes change accordingly [Fig.~\ref{fig:structure}(c)].   Now, both $\sigma^x$ and $\sigma^y$ become nonzero under $\mathcal{M}_1$ and $\mathcal{TM}_2$ operations.  Hence, for the $m'm'm$ group ($\theta= 30^\circ$), we obtain $\bm{\sigma} = [\sigma^x, \sigma^y, 0] = [\sigma_{yz}, \sigma_{zx}, 0]$.  Third, when $\theta$ = 90$^\circ$,  the anomalous Hall vector resorts to the shape \bm{$\sigma$} = $[0, \sigma^y, 0]$ = $[0, \sigma_{zx}, 0]$ because the $\mathcal{M}_1$ and $\mathcal{M}_2$ mirror planes are parallel and perpendicular to the $x$-axis, respectively, as shown in Fig.~\ref{fig:structure}(d).  Finally, the $2'/m'$ group ($\theta$ = 15$^\circ$) only possesses a combined symmetry operation $\mathcal{TM}_3$, where the mirror plane $\mathcal{M}_3$ is parallel to the $x$-$y$ plane and perpendicular to the $z$-axis [Fig.~\ref{fig:structure}(b)].  For this group, the nonzero elements of the AHC are $\bm{\sigma} = [\sigma^x, \sigma^y, 0] = [\sigma_{yz}, \sigma_{zx}, 0]$.  The above results of symmetry analysis can also be obtained by the Neumann principle~\cite{Seemann2015}, wherein all symmetry operations of the corresponding magnetic point group are applied to the conductivity tensor.  Additionally, the cluster multipole theory~\cite{Suzuki_2017} serves as another valuable analysis tool, revealing the shape of conductivity tensor by assessing the cluster multipole moment, which acts as a macroscopic magnetic order.

	\begin{table*}[t]\footnotesize
		\caption{The magnetic space group (MSG), magnetic point group (MPG), and off-diagonal elements of anomalous Hall conductivity for Mn$_3$Sn upon the variation of the spin rotation angle ($\theta$) within the kagome plane.  The presence and absence of off-diagonal elements in the anomalous Hall conductivity are denoted by ``$\checkmark$" and ``$\times$", respectively. }
		\label{tab:group}
		\begin{ruledtabular}
			\begingroup
			\setlength{\tabcolsep}{3.5pt} 
			\renewcommand{\arraystretch}{1.5} 
			\begin{tabular}{lccccccccccccc}
				
				$\theta$   &$0^{\circ}$&$15^{\circ}$&$30^{\circ}$&$45^{\circ}$&$60^{\circ}$&$75^{\circ}$&$90^{\circ}$&$105^{\circ}$&$120^{\circ}$&$135^{\circ}$&$150^{\circ}$&$165^{\circ}$&$180^{\circ}$ \\
				
				\cline{1-14}		
				MSG & $Cmc'm'$ & $P2_1'/m'$ & $Cm'cm'$ & $P2_1'/m'$ & $Cmc'm'$ & $P2_1'/m'$ & $Cm'cm'$ & $P2_1'/m'$ & $Cmc'm'$ & $P2_1'/m'$ & $Cm'cm'$ & $P2_1'/m'$  & $Cmc'm'$\\
				
				MPG & $m'm'm$ & $2'/m'$ & $m'm'm$ & $2'/m'$ & $m'm'm$ & $2'/m'$ & $m'm'm$ & $2'/m'$ & $m'm'm$ & $2'/m'$ & $m'm'm$ & $2'/m'$  & $m'm'm$\\
				
				$\sigma_{xy}$ &$\times$ & $\times$  & $\times$ & $\times$ & $\times$ & $\times$ & $\times$ & $\times$ & $\times$ & $\times$ & $\times$  & $\times$  & $\times$\\		
				$\sigma_{yz}$ & $\checkmark$ & $\checkmark$  & $\checkmark$ & $\checkmark$ & $\checkmark$ & $\checkmark$ & $\times$ & $\checkmark$ & $\checkmark$ & $\checkmark$ & $\checkmark$  & $\checkmark$  & $\checkmark$\\
				
				$\sigma_{zx}$ & $\times$ & $\checkmark$  & $\checkmark$ & $\checkmark$ & $\checkmark$ & $\checkmark$ & $\checkmark$ & $\checkmark$ & $\checkmark$ & $\checkmark$ & $\checkmark$  & $\checkmark$  & $\times$\\
				
			\end{tabular}
			\endgroup
		\end{ruledtabular}
	\end{table*}
	
	\begin{figure}[t]
		\includegraphics[width=1\columnwidth]{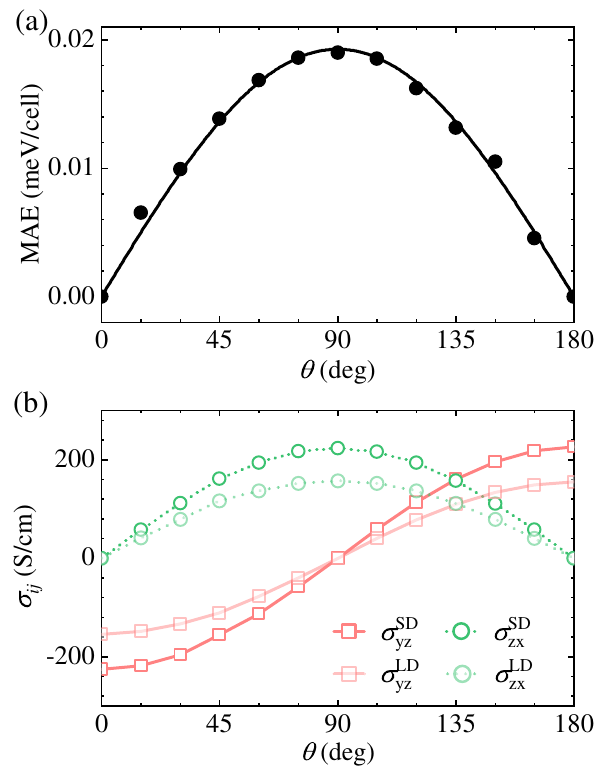}
		\caption{(Color online)  (a) Magnetic anisotropy energy of Mn$_3$Sn as a function of the spin rotation angle $\theta$.  (b) The total AHC ($\sigma_{yz}$ and $\sigma_{zx}$) as a function of $\theta$ calculated using a small disorder (SD) parameter, $\mathcal{V}=2 \ \textnormal{eV}^2a_0^3$, and a large disorder (LD) parameter, $\mathcal{V}=42 \ \textnormal{eV}^2a_0^3$, which correspond to the longitudinal conductivities of $\sigma_{xx} = 1.7\times10^5$ S/cm and $6.4\times10^3$ S/cm, respectively.}
		\label{fig:AHC_angle}
	\end{figure}
	
	In the magnetic group analysis detailed above, we have determined all possible nonzero elements of the AHC vector corresponding to different spin rotation angles, as summarized in Table~\ref{tab:group}. However, one may wonder which angle $\theta$ represents the magnetic ground state of Mn$_3$Sn.  For the noncollinear AFMs considered here, the magnetic anisotropy energy (MAE) can be defined as the total energy difference between distinct spin configurations:
	\begin{equation}\label{eq:MAE}
		\textnormal{MAE}(\theta)=E_{\theta\neq0}-E_{\theta=0}.
	\end{equation}
	The variation of MAE with respect to $\theta$ is illustrated in Fig.~\ref{fig:AHC_angle}(a), from which it becomes evident that the spin configuration with $\theta = 0^\circ$ (magnetic space group $Cmc'm'$) represents the magnetic ground state of Mn$_3$Sn, being 0.019 meV lower in energy than the configuration with $\theta = 90^\circ$ (magnetic space group $Cm'cm'$). This finding is consistent with a prior theoretical calculation~\cite{GuoGY2017}. Figure~\ref{fig:AHC_angle}(a) only presents the MAE results within the $\theta$ range from $0$ to $\pi$, as the 120$^\circ$ noncollinear spin order exhibits a discrete two-fold energy degeneracy, rendering MAE($\theta$) = MAE($\theta+\pi$). This degeneracy characteristic aligns with that observed in the cases of noncollinear antiferromagnetic Mn$_3X$N ($X =$ Ga, Zn, Ag, or Ni)\cite{Zhou2019} as well as two-dimensional van der Waals layered magnets 1$T$-CrTe$_2$~\cite{XX-Yang2021}, Fe$_n$GeTe$_2$ ($n$ = 3, 4, 5)~\cite{XX-Yang2021a}, and Cr$XY$($X$ = S, Se, Te; $Y$ = Cl, Br, I)~\cite{XX-Yang2022}.
	
	Correspondingly, Fig.~\ref{fig:AHC_angle}(b) portrays the total AHC ($\sigma_{ij}$) as a function of $\theta$, computed using two representative disorder parameters. At $\theta = 0^\circ$ or $180^\circ$, solely the $yz$ component of AHC exhibits a nonzero value, whereas at $\theta = 90^\circ$, only the $zx$ component is nonzero. For other $\theta$ values, both $yz$ and $zx$ components contribute to the AHC, harmonizing seamlessly with our magnetic group analysis. The AHC is depicted over the range $0\leq\theta\leq\pi$, while the results for $\pi\leq\theta\leq2\pi$ can be acquired by following the relation $\sigma(\theta)=-\sigma(\theta+\pi)$. This observation arises from the fact that the spin state at $\theta+\pi$ constitutes the time-reversed counterpart of the state at $\theta$, and the AHC maintains an odd symmetry under time-reversal operations~\cite{Zhou2019,XX-Yang2021}. Another intriguing observation from Fig.~\ref{fig:AHC_angle}(b) is that the AHC is enhanced across all spin rotation angles when the disorder parameter is decreased. This phenomenon aligns precisely with the disorder-induced amplification of anomalous transport phenomena previously observed in topological semimetals $M$F$_3$ ($M = $ Mn, Pd)~\cite{XD-Zhou2022}.
	
\begin{figure}[t]
		\includegraphics[width=\columnwidth]{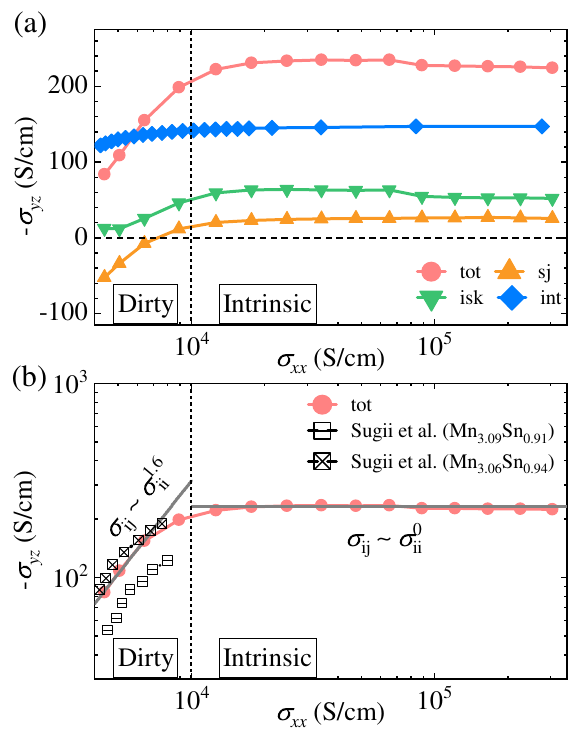}
		\caption{(Color online)  (a) The total AHC ($\sigma_{yz}$) and its partitioning into the intrinsic ($\sigma^\textnormal{int}_{yz}$), side jump ($\sigma^\textnormal{sj}_{yz}$), and skew scattering ($\sigma^\textnormal{isk}_{yz}$) components as a function of the longitudinal conductivity ($\sigma_{xx}$) for the magnetic ground state ($\theta=0^\circ$) of Mn$_3$Sn.  (b) The scaling relation between $\sigma_{yz}$ and $\sigma_{xx}$.  Experimental data from Sugii et al.~\cite{sugii2019anomalous} are provided for comparison.}
		\label{fig:AHC-disorder}
	\end{figure}

	Based on the dependence of the total AHC ($\sigma_{ij}$) on the longitudinal conductivity ($\sigma_{ii}$), three distinct scaling relations have been proposed for various magnetic materials~\cite{Nagaosa2010,YangSY2020,HuangM2021,Miyasato2007}: $\sigma_{ij} \propto \sigma_{ii}^{2}$ or $\sigma_{ii}^{1}$ in the clean regime ($\sigma_{ii} > 10^6$ S/cm), $\sigma_{ij}\propto \sigma_{ii}^0$ in the intrinsic regime ($10^4 < \sigma_{ii} < 10^6$ S/cm), and $\sigma_{ij}\propto \sigma_{ii}^{1.6}$ in the dirty regime ($\sigma_{ii} < 10^4$ S/cm).  While earlier theoretical investigations have primarily focused on the intrinsic AHE in Mn$_3$Sn~\cite{Chen2014,Kubler_2014,Suzuki_2017,GuoGY2017,ZhangYang2017}, the influence of scattering-dependent extrinsic mechanisms has yet to be explored comprehensively.  Figure~\ref{fig:AHC-disorder}(a) showcases the total AHC ($\sigma_{yz}$) and its decomposition ($\sigma^\textnormal{int}_{yz}$, $\sigma^\textnormal{sj}_{yz}$, and $\sigma^\textnormal{isk}_{yz}$) as a function of longitudinal conductivity $\sigma_{xx}$ for Mn$_3$Sn in its magnetic ground state ($\theta=0^\circ$). Notably, Mn$_3$Sn predominantly lies within the dirty and intrinsic regimes due to its $\sigma_{xx}< 10^6$ S/cm. As $\sigma_{xx}$ increases, the total AHC $\sigma_{yz}$ rises and gradually approaches a constant plateau of $-230$ S/cm for $\sigma_{xx} >10^4$ S/cm. In the intrinsic regime, the intrinsic AHC $\sigma^\textnormal{int}_{yz}$ ($\approx -146$ S/cm) plays a dominant role, aligning well with previous theoretical calculations~\cite{Kubler_2014,GuoGY2017,Suzuki_2017,ZhangYang2017}. Meanwhile, the extrinsic mechanisms ($\sigma^\textnormal{sj}_{yz} + \sigma^\textnormal{isk}_{yz}$) assume a secondary role, contributing to 34\% of the total AHC.  Given that the intrinsic mechanism is independent of scattering, it should be much less affected by changes in longitudinal conductivity, as compared to extrinsic mechanisms.  
	
	Our calculations indeed demonstrate that the intrinsic contribution remains fairly stable within the dirty regime ($\sigma_{xx}< 10^4$ S/cm). Conversely, skew scattering rapidly diminishes towards zero, while side jump experiences a significant increase, primarily governing the declining trend of the total AHC.  A recent experimental study has reported a reduction in the total AHC as $\sigma_{xx}$ decreases below $10^4$ S/cm~\cite{sugii2019anomalous}. However, this work~\cite{sugii2019anomalous} mentioned that the contribution of the side jump mechanism can be ruled out due to the weak spin-orbit coupling strength of Mn 3$d$ electrons, suggesting that the reduction in total AHC is driven by the intrinsic mechanism.  In direct comparison, Fig.~\ref{fig:AHC-disorder}(b) demonstrates the excellent agreement between our calculations and the experimental results~\cite{sugii2019anomalous}.  In the dirty regime ($\sigma_{xx} < 10^4$ S/cm), a scaling relation of $\sigma_{yz} \sim \sigma_{xx}^{1.6}$ is evident, highlighting the pronounced significance of the extrinsic side jump mechanism. Consequently, our analysis tends to a conclusion that whereas the intrinsic mechanism dominates in the intrinsic regime, the large reduction of AHC in dirty regime is primarily attributed to the contribution of the side jump mechanism.
	
	\begin{figure*}
		\includegraphics[width=2\columnwidth]{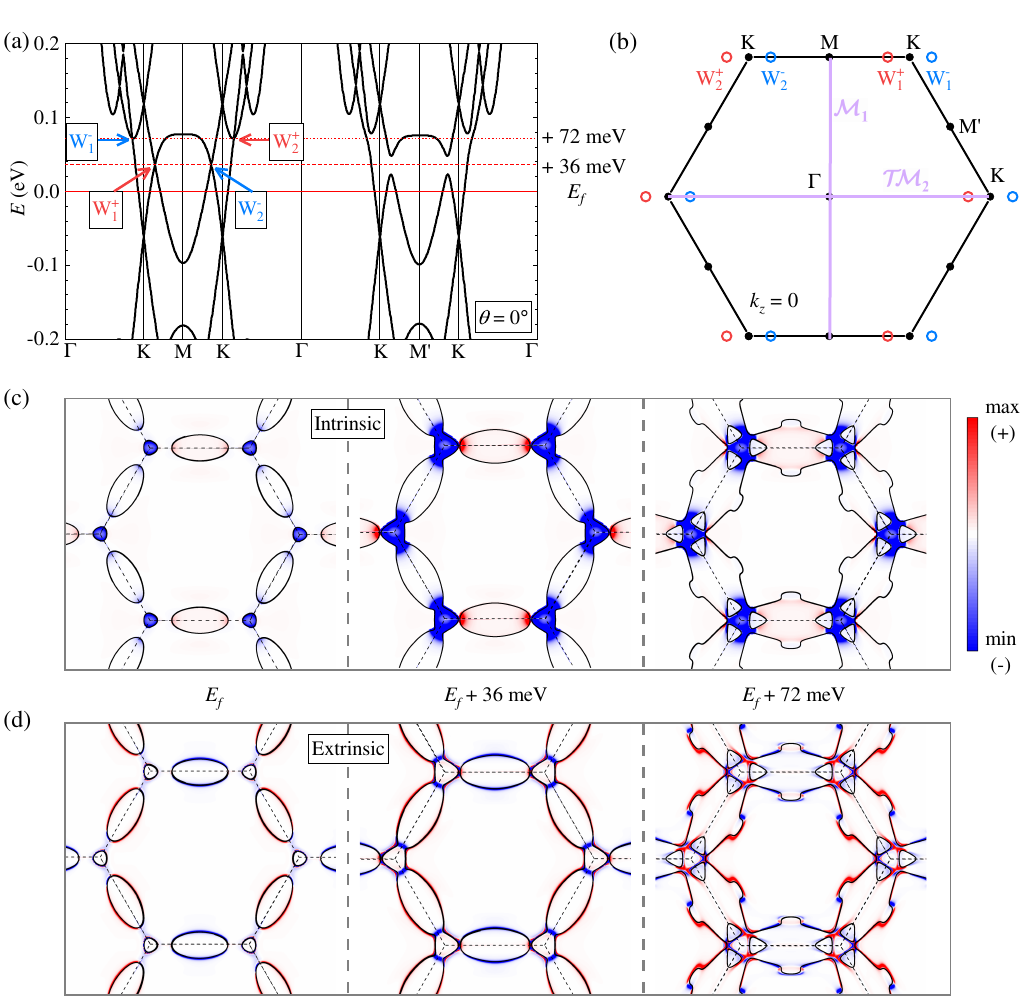}
		\caption{(Color online)  (a) The band structure of Mn$_3$Sn in its magnetic ground state ($\theta = 0^\circ$). The energy positions of Weyl points (W$_1^\pm$ and W$_2^\pm$) are indicated by red dashed lines. (b) The distribution of Weyl points on the $k_z= 0$ plane in momentum space. Red and blue circles denote Weyl points with different chiralities. Purple lines represent $\mathcal{M}_1$ and $\mathcal{TM}_2$ symmetries. (c,d) Momentum-resolved intrinsic and extrinsic AHC $\sigma_{yz}$ (color maps) as well as Fermi surfaces (black lines) on the $k_z= 0$ plane for three distinct Fermi energies ($E_f$ in the left panels, $E_f+36$ meV in the middle panels, $E_f+72$ meV in the right panels). The disorder parameter is chosen to be $\mathcal{V}=2 \ \textnormal{eV}^2a_0^3$, at where the longitudinal conductivity $\sigma_{xx} = 1.7\times10^5$ S/cm.}
		\label{fig:band_berry}
	\end{figure*}
	
	The intrinsic mechanism of AHC stems from the presence of nonvanishing Berry curvature in momentum space and can be largely enhanced by topological features in the band structure like Weyl nodal points.  Nevertheless, the link between extrinsic mechanisms of AHC and the underlying band structure remains less elucidated.  Recent theoretical and experimental investigations have illuminated the existence of magnetic Weyl fermions near the Fermi energy ($E_f$) in Mn$_3$Sn~\cite{Muhammad2017,Kuroda2017,ZhangYang2017,Yang_2017}. As these Weyl points can be interpreted as effective magnetic monopoles in momentum space, the increased Berry curvature in proximity to these points contributes to a substantially amplified AHC.  The band structure calculated with spin-orbit coupling for the ground state spin configuration ($\theta=0^\circ$) of Mn$_3$Sn is presented in Fig.~\ref{fig:band_berry}(a).  The time-reversal symmetry breaking triggers the emergence of multiple pairs of Weyl points at varying energy levels.  For our analysis, we focus on those near $E_f$ as they are pertinent to the anomalous transport properties.  Owing to the $\mathcal{M}_1$ and $\mathcal{TM}_2$ symmetries within the $m'm'm$ group, all K points in the first Brillouin zone are equivalent, while two inequivalent M points are labeled M and M$'$.  In the vicinity of the M point, an intersection between a parabolic band and an anti-parabolic-like band engenders Weyl points ($\rm W^+_1, W^-_2$) at $E_f + 36$ meV, accompanied by their counterparts ($\rm W^-_1, W^+_2$) at $E_f + 72$ meV. However, no Weyl points are present near the M$'$ point. The spatial distribution of Weyl points along the K-M-K path on the $k_z = 0$ plane is showcased in Fig.~\ref{fig:band_berry}(b).  Upon shifting the Fermi energy upwards to 36 and 72 meV, the intrinsic AHC around the Weyl points exhibits a sharp increase, as depicted in Fig.~\ref{fig:band_berry}(c). This observation affirms the inherent enhancement of intrinsic AHC through topological Weyl nodal structures.  Furthermore, Fig.~\ref{fig:band_berry}(d) illustrates the extrinsic AHC at the same Fermi energies. In contrast to the intrinsic AHC, the extrinsic AHC is primarily distributed along the Fermi surface sheets, indicating a more substantial contribution from the Fermi surfaces as compared to the Fermi sea.
	
	\begin{figure*}[t]
		\includegraphics[width=2\columnwidth]{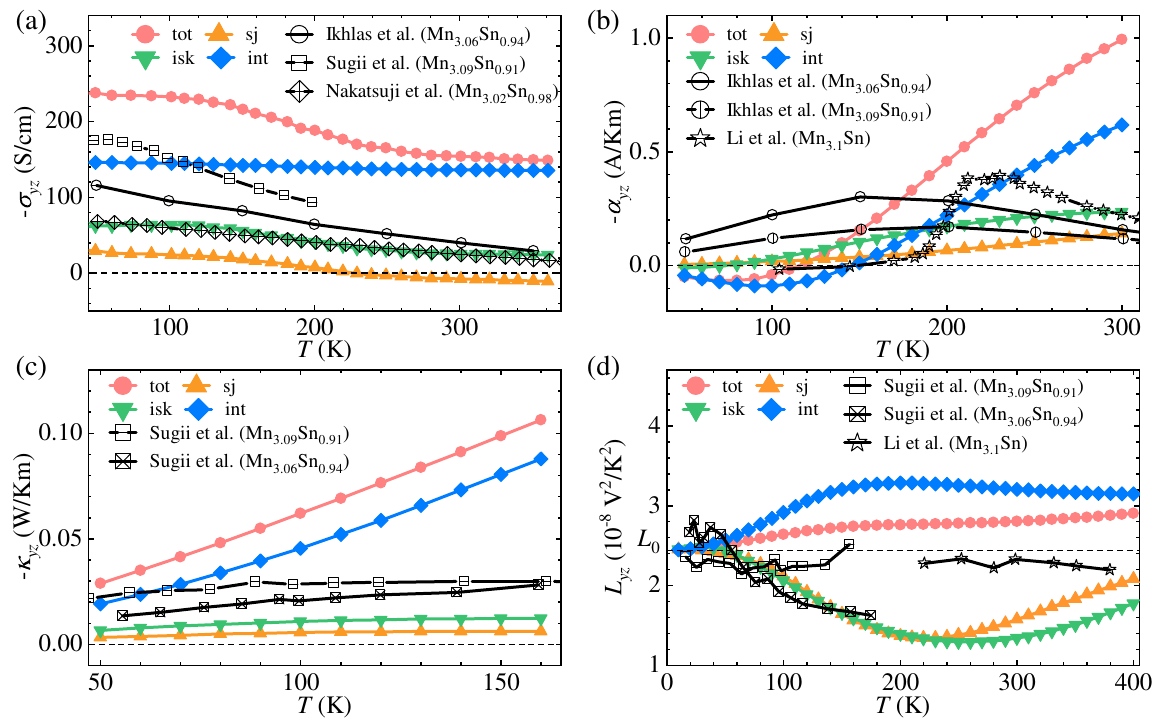}
		\caption{(Color online)  (a) The temperature-dependent AHC in comparison  to experimental data from Nakatsuji~\cite{Nakatsuji2015}, and Ikhlas~\cite{Muhammad2017}, Sugii~\cite{sugii2019anomalous} et al.  (b) The temperature-dependent ANC  in comparison to experimental data from Ikhlas~\cite{Muhammad2017} and Li~\cite{LiXK2017} et al.  (c) The temperature-dependent ATHC in comparison to experimental data from Sugii~\cite{sugii2019anomalous} et al.  (d) The anomalous Lorenz ratio in comparison to experimental data from Li~\cite{LiXK2017} and Sugii~\cite{sugii2019anomalous} et al.  The horizontal dashed line in (d) marks the free-electron Lorenz number $L_0 = 2.44 \times 10^{-8}\ \rm V^2/K^2$.  The disorder parameter is selected as $\mathcal{V} = 1.87$ eV$^2$a$_0^3$, where the longitudinal conductivity $\sigma_{xx} = 1.66 \times 10^5$ S/cm.}
		\label{fig:AHC-temperature}
	\end{figure*}

	Next, we turn to the variation in the anomalous transport properties of Mn$_3$Sn with temperature.  By utilizing the experimentally derived relationship between the longitudinal conductivity $\sigma_{ii}$ and temperature, we can easily map our results of $\sigma_{ij}$ to temperature.  Figure~\ref{fig:AHC-temperature}(a) illustrates the calculated AHC as a function of temperature, compared with available experimental data. Notably, three previous experimental studies~\cite{Muhammad2017,Nakatsuji2015,sugii2019anomalous} show noticeable discrepancies in the magnitude of the AHC. This can be attributed to variations in the chemical composition of the samples, as the Mn-to-Sn atomic ratio deviates from the ideal 3:1, leading to differences in Mn content across different samples. However, the crucial observation is that all of these studies consistently demonstrate a declining trend in AHC with increasing temperature. 

	we observe that the intrinsic AHC ($\sigma^\textnormal{int}_{yz}$) remains relatively constant over the entire temperature range, and the reduction in AHC with increasing temperature can be attributed to the behavior of the side jump and skew scattering ($\sigma^\textnormal{sj}_{yz}+\sigma^\textnormal{isk}_{yz}$) contributions.  This underlines the significant role that the extrinsic mechanisms play in shaping the AHE of Mn$_3$Sn at higher temperatures.  We have to remark that with increasing temperature, the spin fluctuation away from the equilibrium direction $-$ which are not included into our analysis explicitly $-$ may become more prominent. This will eventually result in a suppression of the averaged over possible angles values of the AHC [Fig.~\ref{fig:AHC_angle}(b)]. The presence of this mechanism may explain the stronger decay of the AHC with $T$ observed experimentally. Since the latter effective averaging also results in smoother behavior of the AHC with band filling, similar effect of suppression at higher temperatures can be experienced in the case of ANC, discussed below.
	
	The total ANC ($\alpha_{yz}$) along with its components ($\alpha^\textnormal{int}_{yz}$, $\alpha^\textnormal{sj}_{yz}$, and $\alpha^\textnormal{isk}_{yz}$) are computed using Eq.~\eqref{eq:ANC} and presented as a function of temperature in Fig.~\ref{fig:AHC-temperature}(b). It can be observed that the ANC gradually increases with rising temperature, with a particularly pronounced rise occurring when the temperature exceeds 100 K. This upward trend contrasts with experimental findings, which have demonstrated a decrease in ANC as the temperature goes beyond 150 K~\cite{Muhammad2017} or 200 K~\cite{LiXK2017}. Notably, it is worth mentioning that a phase transition from a noncollinear antiferromagnetic structure to a helical spin structure has been reported around 200 K~\cite{LiXK2017}. In our calculations, we have exclusively considered a perfect magnetic crystal featuring a 120$^\circ$ noncollinear antiferromagnetic structure, thereby excluding the influence of any additional phase transitions.  Furthermore, in addtion to the mechanism discussed above for the case of AHC, as the temperature increases, phonon thermal dynamics becomes more pronounced, which undoubtedly impacts the anomalous transport properties of magnetic materials. The Gaussian disorder model~\cite{Czaja2014} employed in our work broadly encompasses all ``mean-field" scattering channels. However, the intricate details of electron scattering arising from phonons are not explicitly accounted for. Consequently, as temperature rises, our calculated ANC is expected to increase due to its positive temperature-dependent nature.
	
	Subsequently, we delve into the ATHE of Mn$_3$Sn, akin to the thermal counterpart of AHE, as illustrated in Fig.~\ref{fig:AHC-temperature}(c).  At lower temperatures (below 100 K), the calculated ATHC is in good agreement with experimental results\cite{sugii2019anomalous}. As the temperature increases, the intrinsic ATHC ($\kappa^\textnormal{int}_{yz}$) exhibits a monotonic increase, leading to an overall rising trend in the total ATHC ($\kappa_{yz}$). However, experimental observations have indicated a relatively minor temperature dependence in ATHC~\cite{sugii2019anomalous}, fitting to the behaviors of our calculated extrinsic ATHC ($\kappa^\textnormal{isk}_{yz} + \kappa^\textnormal{sj}_{yz}$).  The anomalous thermal and electrical transports can be interconnected through the anomalous Lorenz ratio, defined as
	\begin{equation}\label{eq:L}
		L_{ij} = \kappa_{ij} / (\sigma_{ij}T),
	\end{equation}
	which has be employed to judge the contributions of intrinsic and scattering to the AHE~\cite{Onose_Lor2008,Shiomi_Lor2009,Shiomi_Lor2010}. Similar to the AHC, ANC, and ATHC, $L_{ij}$ can also be decomposed into three parts:
	\begin{equation}\label{eq:L-decomp}
		L_{ij} = L_{ij}^\textnormal{int} + L_{ij}^\textnormal{sj} + L_{ij}^\textnormal{isk},
	\end{equation}
	where $L_{ij}^\textnormal{int}$ represents the intrinsic contribution, $L_{ij}^\textnormal{sj}$ and $L_{ij}^\textnormal{isk}$ are the extrinsic contributions from side jump and skew scattering, respectively.  As the temperature approaches to zero, $L_{ij}$ converges to the free-electron Lorenz number, commonly referred to as the Wiedemann-Franz law:
	\begin{equation}\label{eq:L0}
		L_{ij} (T\rightarrow0) \approx L_0 = \dfrac{\pi^2k_B^2}{3e^2} = 2.44\times10^{-8}\ \textnormal{V}^2/\textnormal{K}^2.
	\end{equation}
	Examining Fig.~\ref{fig:AHC-temperature}(d), we observe that when the temperature is less than 50 K, the calculated $L_{yz}$ closely aligns with $L_0$. This indicates that at low temperatures, the intrinsic mechanism predominantly governs the anomalous electrical and thermal transport in Mn$_3$Sn, suggesting that transverse charge and heat currents flow in a nearly dissipationless way. As temperature rises, the transverse heat current carried by conducting electrons is expected to experience progressively increased dissipation due to inelastic scattering with phonons. For instance, above 50 K, experimentally measured $L_{yz}$ for Mn$_{3.06}$Sn$_{0.94}$\cite{sugii2019anomalous} deviates noticeably from $L_0$, signaling a crossover in the dominant role from intrinsic mechanism to extrinsic mechanisms. This is consistent with our calculations of extrinsic contributions ($L^\textnormal{sj}_{yz}+L^\textnormal{isk}_{yz}$) to the anomalous Lorenz ratio as the temperature increases. However, for the Mn$_{3.09}$Sn$_{0.91}$ sample~\cite{sugii2019anomalous} and another experimental study of Mn$_{3}$Sn~\cite{LiXK2017}, the deviation of $L_{yz}$ from $L_0$ is not substantial, indicating the potential persistence of the intrinsic mechanism's predominance over the extrinsic ones. Thus, in our calculations, the anomalous Lorenz ratio $L_{yz}$ remains relatively close to $L_0$ across the entire temperature range.

	\section{Summary}
	In summary, we have systematically studied the intrinsic and extrinsic anomalous Hall, anomalous Nernst, and anomalous thermal Hall effects in noncollinear antiferromagnetic Mn$_3$Sn, utilizing advanced first-principle calculations and magnetic group analysis. In our study, the intrinsic contribution is associated with the Berry phase effect of relativistic bands within a pristine crystal, free from impurities. All additional contributions arising from scatterings on impurities or disorder are classified as extrinsic. The spin-independent impurity potential utilized in our study can be understood as representing nonmagnetic impurities in magnetic materials. With the incorporation of spin-orbit coupling, the electron's spin becomes intricately dependent on the modification of its orbital angular momentum during scattering. Consequently, the transverse flow of spin-polarized electrons induced by scattering on nonmagnetic impurities is indeed feasible.  The definitions of intrinsic and extrinsic contributions to the anomalous transport properties align with established conventions in the majority of previous research. 
	
	We first identified the nonvanishing tensor elements of anomalous Hall conductivity for diverse coplanar noncollinear spin configurations, according to symmetry requirements under relevant magnetic point groups.  Upon the collective rotation of all spins within the kagome plane, the anomalous Hall conductivity showcases periodic patterns, giving rise to a pronounced magnetic anisotropy.  Previous theoretical works have primarily focused on studying the intrinsic anomalous transport properties of Mn$_3$Sn through Berry curvature calculations, with relatively less attention given to extrinsic mechanisms. Through computations of the total anomalous Hall conductivity and its constituent components, we have unveiled that the intrinsic mechanism uniformly dominates within the intrinsic regime, especially when the longitudinal conductivity $\sigma_{xx}$ surpasses $10^4$ S/cm.  The intrinsic mechanism can be traced to substantial Berry curvatures encircling Weyl points proximate to the Fermi energy.  In the realm of the dirty regime ($\sigma_{xx}<10^4$ S/cm), extrinsic mechanisms, notably the side jump, emerge as potent contributors, which brings our theoretical results closer to experimental measurements.  This extrinsic anomalous Hall conductivity predominantly stems from electronic states positioned precisely at the Fermi surfaces.  Moreover, our findings with regards to the anomalous thermal Hall effect and anomalous Lorenz ratio compare well with experimental outcomes at low temperatures, consistently indicating the dominated role of the intrinsic mechanism.  As the temperature rises, a certain degree of deviation between theoretical and experimental results becomes apparent. These deviations may be attributed to enhanced phonon scattering and increased complexity in the internal structure of the crystal, factors that are not fully accounted for in the Gaussian disorder model.  Through these comprehensive insights, our study has substantially enriched the understanding of anomalous transport phenomena in noncollinear antiferromagnetic Mn$_3$Sn. Furthermore, our work offers valuable perspectives for potential applications in the realms of spintronics and spin caloritronics, harnessing the distinctive attributes of noncollinear antiferromagnetic materials.
	
	\begin{acknowledgments}
		This work is supported by the National Key R\&D Program of China (Grants No. 2022YFA1403800, No. 2022YFA1402600, and No. 2020YFA0308800), the National Natural Science Foundation of China (Grants No. 12274027 and No. 11874085), the Science \& Technology Innovation Program of Beijing Institute of Technology (Grant No. 2021CX01020).  Y.M. acknowledges the Deutsche Forschungsgemeinschaft (DFG, German Research Foundation) TRR 288—422213477 (Project No. B06). Y.M., W.F., and Y.Y. acknowledge the funding under the Joint Sino-German Research Projects (Chinese Grant No. 12061131002 and German Grant No. 1731/10-1) and the Sino-German Mobility Program (Grant No. M-0142).
		
	\end{acknowledgments}
	
	\bibliography{references}

\end{document}